# *PROTEUS*: Rule-Based Self-Adaptation in Photonic NoCs for Loss-Aware Co-Management of Laser Power and Performance

Sairam Sri Vatsavai
Electrical and Computer Engineering
University of Kentucky
Lexington, USA
sairam_srivatsavai@uky.edu

Venkata Sai Praneeth Karempudi
Electrical and Computer Engineering
University of Kentucky
Lexington, USA
kvspraneeth@uky.edu

Ishan Thakkar
Electrical and Computer Engineering
University of Kentucky
Lexington, USA
igthakkar@uky.edu

*Abstract*— The performance of on-chip communication in the state-of-the-art multi-core processors that use the traditional electronic NoCs has already become severely energy-constrained. To that end, emerging photonic NoCs (PNoC) are seen as a potential solution to improve the energy-efficiency (performance per watt) of on-chip communication. However, existing PNoC designs cannot realize their full potential due to their excessive laser power consumption. Prior works that attempt to improve laser power efficiency in PNoCs do not consider all key factors that affect the laser power requirement of PNoCs. Therefore, they cannot yield the desired balance between the reduction in laser power, achieved performance and energy-efficiency in PNoCs. In this paper, we present *PROTEUS* framework that employs rule-based self-adaptation in PNoCs. Our approach not only reduces the laser power consumption, but also minimizes the average packet latency by opportunistically increasing the communication data rate in PNoCs, and thus, yields the desired balance between the laser power reduction, performance, and energy-efficiency in PNoCs. Our evaluation with PARSEC benchmarks shows that our *PROTEUS* framework can achieve up to 24.5% less laser power consumption, up to 31% less average packet latency, and up to 20% less energy-per-bit, compared to another laser power management technique from prior work.

## I. INTRODUCTION

To support the increasing demand for on-chip data communication in modern multicore processors, the use of electrical networks-on-chip (ENoCs) has become a norm. However, the performance of the state-of-the-art ENoCs is projected to scale poorly for the emerging data-centric applications (e.g., internet-of-things (IoT) related applications), primarily due to the energy-constrained bandwidth of ENoCs. To this end, with the recent advancements in silicon photonics, photonic networks-on-chip (PNoCs) are being considered as potential replacements for ENoCs. This is because PNoCs can provide several advantages over ENoCs, such as distance-independent higher datarates and lower dynamic energy consumption. However, the state-of-the-art PNoC architectures (e.g., [17], [18]) require a non-trivial amount of optical power from their laser source, mainly because of the high insertion loss of photonic devices in their constituent photonic links [38]. The high laser power overheads can offset the high aggregated data-rate and energy-efficiency advantages of PNoCs. Therefore, it is imperative to innovate new techniques that can reduce the optical power consumption in future PNoC architectures.

Several techniques have been proposed in prior works (e.g., [1]-[9]) that aim to reduce the laser power consumption in PNoCs. Some of these techniques dynamically adjust the optical power extracted from the off-chip laser sources, in response to either the temporal and spatial variations in the network traffic (e.g., [1]-[5]) or the change in the insertion loss for every photonic data packet transfer (e.g., [6]-[7]). In addition, recent works [8] and [9] take a holistic approach and use machine learning predictors for leveraging the variations in both the network traffic and insertion loss, to achieve greater savings in laser power consumption. However, all these techniques can incur dauntingly high overheads for dynamic monitoring of the network traffic (e.g., in [1]-[5]), integration of costly on-chip optical amplifiers (e.g., in [7]), runtime execution of NP-hard optimization heuristics (e.g., in [6]), or runtime inference of the machine learning models (e.g., in [8], [9]). Moreover, these techniques do not consider the effects of bit-error rate (BER) penalty due to various sources of errors (e.g., crosstalk) on the laser power utilization and performance of photonic links and PNoCs. As a result, these techniques are not able to achieve the required practical balance between the reduction in laser power and achieved performance in PNoCs. To achieve such power-performance balance in PNoCs, recent works [30] and [31] employ data approximation techniques to opportunistically trade the communication reliability for reduced laser power and/or improved performance in PNoCs. However, to gain substantial benefits, these techniques require the accuracy or reliability goals of target applications to be relaxed, which can be achieved only for a select few inherently error-tolerant applications. This constraint limits the applicability of such techniques.

In contrast to these dynamic techniques from prior work, we advocate for a hybrid (static + dynamic) solution in this paper as part of our proposed *PROTEUS* framework. Instead of dynamically tuning the optical power extracted from the laser source, our *PROTEUS* framework statically minimizes the required optical power extraction from the laser source at the design-time, by optimizing two key photonic link configuration parameters to minimize the BER power penalty in PNoCs without reducing the reliability of communication. Then, at the runtime, *PROTEUS* dynamically adapts the photonic link configuration in response to the changing insertion loss for every photonic packet transfer, to achieve and maintain the balance between the reduction in laser power and achieved performance. For dynamic adaptation, *PROTEUS* relies on simple rules that are derived from an offline search heuristic. *PROTEUS* stores these rules in lookup tables to enable their easy reference during the runtime of PNoCs. Our novel contributions in this paper are summarized below:

- We present a design-time technique that minimizes the crosstalk related BER power penalty in PNoCs by optimizing two key photonic link configuration parameters, to ultimately reduce the requirement of laser power in PNoCs;
- We present light-weight techniques for implementing self-adaptation of photonic link configuration, and provide detailed overhead analysis of these techniques;
- We integrate these design-time optimization and runtime self-adaptation techniques into a holistic framework called *PROTEUS*, to achieve a loss-aware balance between the laser power consumption and performance of PNoCs;

---

[1] This work was supported by a seed grant from the University of Kentucky.

- We evaluate *PROTEUS* by implementing it on a well-known PNoC architecture and compare it with other laser power management techniques from prior works [3] and [7].

## II. FUNDAMENTALS OF PHOTONIC NoCs (PNoCs)

### A. Physical-Layer Architecture and Operation of PNoCs

In this subsection, we explain the physical-layer design and operation of PNoC architectures. We use the crossbar-based PNoC from [25] as an example PNoC architecture in this paper, the physical-layer layout of which is illustrated in Fig. 1. The PNoC in Fig. 1 consists of serpentine links as its building blocks. Every such link in the PNoC consists of one or more photonic waveguides spanning the PNoC chip, depending on the specific variant of the physical-layer architecture [25]. In this paper, we consider one photonic waveguide per link. Every such single-waveguide photonic link in the PNoC connects multiple gateway interfaces (GIs) with one another. A GI connects to multiple parallelly laid-out photonic links, and interfaces a cluster of processing cores (e.g., a cluster of four cores in Fig. 1) with the links. Typically, out of all the GIs that are connected to a single link, some GIs can write photonic data into the link and the others can read photonic data from the link, to enable the multiple-writer-multiple-reader (MWMR) type of crossbar configuration [17] in the PNoC.

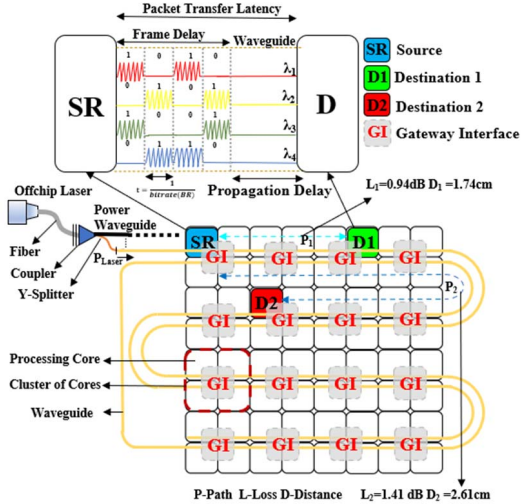

**Fig. 1:** Schematic physical-layer layout of the PNoC architecture from [25]. This figure also explains the concepts of packet frame delay and changing insertion loss ($IL^{dB}$) for every packet transfer.

Each of the photonic links in the PNoC receives some amount of multi-wavelength optical power from an on-chip power waveguide via a power splitter. The power waveguide receives the multi-wavelength optical power from an off-chip laser source via an optical coupler. The input multi-wavelength optical power traverses the individual links to all individual GIs on the chip. Each GI can utilize the wavelengths of input light as parallel dense-wavelength-division-multiplexed (DWDM) carriers for data signals, to enable data communication with one or more other GIs. At a sender GI of the PNoC, every incoming data packet from the source processing core is converted into multiple parallel electrical data signals (a signal is defined here as a sequence of '1's and '0's), which are then modulated onto the DWDM carriers using a bank of modulator MRs (not shown in Fig. 1) to convert them into parallel photonic data signals. These DWDM data signals constitute a photonic data packet that traverses a single-waveguide link to a receiver GI. At the receiver GI, a set of MR filters drops the constituent photonic signals of the photonic data packet onto the adjacent photodetectors, to regenerate the electrical data signals, and consequently, the electrical data packet. This regenerated electrical data packet is then passed on to the destination processing core. Thus, in a PNoC, every data packet is transferred as multiple DWDM data signals.

The transfer of every photonic data packet as multiple DWDM data signals (referred to as $N_\lambda$) in the PNoC enables wrapping of the data packet (packet size is referred to as *PS*) into a short timeframe. This packet timeframe (i.e., $\{|(PS/N_\lambda)|\times(1/BR)\}$, where *BR* is signal bitrate) is referred to as frame delay in Fig. 1. This frame delay, when added to the waveguide propagation delay (Fig. 1), constitutes the latency of transferring the packet between the sender and receiver GIs. It can be reasoned that this transfer latency for a fixed size of the data packet can be reduced by decreasing the packet frame delay (i.e., $\{|(PS/N_\lambda)|\times(1/BR)\}$), which in turn can be achieved in three different ways: *(i)* by increasing $N_\lambda$ in the waveguide, *(ii)* by increasing the bitrate (i.e., *BR*) of each data signal, and *(iii)* by increasing both $N_\lambda$ and *BR*. Each of these three ways can enable wrapping of the data packet into a shorter timeframe, to reduce the packet frame delay, and hence, the packet transfer latency. However, increasing $N_\lambda$ and/or *BR* requires judicious consideration of the inherent tradeoffs among the achievable performance, reliability, and required optical power in the PNoC. Failing to do so can lead to significantly harmed optical power efficiency or nonviable operation of the photonic links and PNoC, as discussed next.

### B. Power-Reliability-Performance Tradeoffs in PNoCs

Designing a photonic link of a PNoC is subject to inherent tradeoffs among the achievable performance (aggregated datarate ($N_\lambda \times BR$), and hence, frame delay ($\{|(PS/N_\lambda)|\times(1/BR)\}$, where *PS* is packet size), required optical power, and reliability [34]. Optimizing these design tradeoffs often involves finding the sweet spot that balances the link's aggregated datarate and power-reliability behavior [34]. The tenacity of this balance depends on how efficiently the provisioned optical power from the off-chip laser source is utilized. The utilization of the provisioned laser power in the link is governed by four different factors, which are formulated in Eq. (1).

$$P_{Max} \geq P_{Laser} \geq IL^{dB} + PP^{dB}\{N_\lambda, BR, Q\} + 10log_{10}(N_\lambda) + S\{BR\} \quad (1)$$

Here, $P_{Laser}$ is the provisioned optical power (in dBm) in the link from the power-waveguide splitter (Fig. 1), the utilization of which depends on the following four factors, as evident from Eq. (1): *(i)* total insertion loss $IL^{dB}$ in dB faced by a single photonic signal in the link, which includes the total propagation and bending loss in the link's waveguide and the total insertion loss of the MR modulators, MR filters, splitters, and couplers; *(ii)* total bit-error rate (BER) power penalty $PP^{dB}$, which is defined as the required increase in the provisioned optical power of a photonic signal to compensate for the reduced bit-error rate (BER) due to various signal degradation phenomena, including inter-modulation crosstalk and inter-signal crosstalk at filter MRs [23]; *(iii)* number of DWDM data signals $N_\lambda$ per waveguide; and *(iv)* the photodetector sensitivity *S* which is a function of BR, which gives the minimum required power of a photonic data signal at the photodetector for the error-free detection of the signal. In addition, the peak value of $P_{Laser}$ in a link should be less then $P_{Max}$ (Eq. (1)), where $P_{Max}$ gives the optical nonlinearity limited maximum allowable optical power in the waveguide (typically, $P_{Max}$ = 20dBm [15], [34]). Thus, $P_{Laser}$ in the link should be not only greater than or equal to the sum of the optical power requirements of all the above four factors, but also less than or equal to $P_{Max}$.

From [34], *S* depends on the *BR* of the photonic signal. Similarly, from [13] and [23], the total power penalty $PP^{dB}$ of a link, as well as the insertion loss values for the modulator and filter MRs (which are part of the total $IL^{dB}$ value for the link), also depend on *BR*. In addition, $PP^{dB}$ of a link also depends on various link configuration parameters, such as quality factor (*Q*) of the MRs, free spectral range (FSR), and wavelength spacing between the adjacent photonic signals in the link [38]. The parameters wavelength spacing and FSR have limited design flexibility due to the limitations imposed by the utilized devices and fabrication technology [34]. For instance, commonly used comb laser sources typically produce output wavelengths with precisely fixed spacings [29], and require additional area-consuming interleavers (e.g., [32]) to provide limited flexibility for tuning their output wavelength spacings. Along the same lines, the state-of-the-art CMOS-compatible MR fabrication technology limits the maximum achievable FSR to 20nm (e.g., [36]). Because of these reasons, for the system-level design of PNoCs, the values of parameters FSR and wavelength spacing can be assumed to be fixed, and consequently, $PP^{dB}$ can be optimized as the function of *BR* and *Q* of MRs (see Section II-C). *As a result, the required $P_{Laser}$ and its utilization in the photonic link ultimately depends on the link configuration parameters $N_\lambda$, Q, and BR.* Thus, for the given value of $IL^{dB}$ in the link, only a finite set of unique values of the ($N_\lambda$, *BR*, *Q*) triplet can satisfy the condition for $P_{Laser}$ given in Eq. (1). From [11] and [15], out of all such values of triplet ($N_\lambda$, *BR*, *Q*), only one triplet value can optimally balance the inherent tradeoffs among the aggregated datarate ($N_\lambda \times BR$), frame delay (packet size/($N_\lambda \times BR$)), and optical power efficiency ($P_{Laser}/(N_\lambda \times BR)$). Thus, any injudicious attempt to increase $N_\lambda$ for improving the packet frame delay can lead to an increased $P_{Laser}$ value, which in turn can result not only in a decreased optical power efficiency ($P_{Laser}/(N_\lambda \times BR)$), but also in a nonviable $P_{Laser}$ value that is greater than $P_{Max}$.

### C. Modeling of $PP^{dB}$ and $IL^{dB}$ as Functions of BR and Q

From [13], Eq. (2) below gives the formula for $PP^{dB}$ (from Eq. (1)) for a photonic signal as the sum of the modulator crosstalk penalty ($PP^{Mod}_{Xtalk}$), filter crosstalk penalty ($PP^{Fil}_{Xtalk}$), and power penalty due to the finite Extinction ratio (ER) of modulation (i.e., the first term in Eq. (2)). From [13], $PP^{Mod}_{Xtalk}$ for a signal does not depend on its *BR*, and for a moderate wavelength spacing of greater than 0.3nm (as assumed for this work), it can be limited below 1dB. *Therefore, we take the fixed 1dB value of $PP^{Mod}_{Xtalk}$ in this paper.* On the other hand, $PP^{Fil}_{Xtalk}$ at a filter MR can be evaluated using Eq. (3) and Eq. (4) given below [13].

$$PP^{dB} = -10\log_{10}\left(\frac{r-1}{r+1}\right) + PP^{Mod}_{Xtalk} + PP^{Fil}_{Xtalk} \quad (2)$$

$$PP^{Fil}_{Xtalk} \approx -10\log_{10}\left(1 - 2\sum_{i=1}^{N_\lambda}\sqrt{\gamma_i}\right) \quad (3)$$

$$\gamma_i = \frac{1}{1+\beta^2} - \frac{1}{2\pi v}Re\left(\frac{1-\exp(-2\pi v(1-j\beta))}{(1-j\beta)^2}\right) \quad (4)$$

Here, *r* extinction power ratio, $\gamma_i$ is the crosstalk power ratio at the filter MR from the $i^{th}$ signal of total $N_\lambda$ signals, $v = f_0/(2Qr_b)$, $\beta = 2Qf_\Delta/f_0$, with *Q* = MR *Q*, $r_b$ = *BR* of the $i^{th}$ signal, $f_0$ is resonance frequency of the MR filter, and $f_\Delta$ denotes the frequency detuning between the $i^{th}$ signal and $f_0$. Fig. 2 gives the modeled $PP^{Fil}_{Xtalk}$ values as a function of *BR* and *Q*. From the figure, for a given value of *BR*, only a unique value of *Q* (as indicated by the optimal curve) can minimize $PP^{Fil}_{Xtalk}$.

In addition, from [13], the insertion losses of MR modulators and filters can be modeled to depend on their *Q* using the Lorentzian shaped transfer function of MRs. The inclusion of these *Q*-dependent insertion loss values of MR modulators and filters in the total $IL^{dB}$ value for the link makes $IL^{dB}$ to depend on the MRs' *Q* as well. *Thus, only a unique combination of Q and BR can minimize both $PP^{dB}$ and $IL^{dB}$ for a photonic link.*

### D. Variation in $IL^{dB}$ for Every Photonic Packet Transfer

In a PNoC, different photonic data packets face different values of the insertion loss $IL^{dB}$. This is because different photonic packets traverse different distances between their sender and receiver GIs. For example, in Fig. 1, a source processing core is highlighted as SR. The photonic packets from SR traverse paths P1 and P2, respectively, to the destination processing cores D1 and D2. Based on the physical layout of the PNoC shown in Fig. 1 on a 2cm×2cm photonic chip for the 22nm technology node, the lengths of paths P1 and P2 are 1.74cm and 2.61cm respectively. Moreover, path P2 also has a waveguide bend. Therefore, based on the various loss model values from Table 1, paths P1 and P2 incur waveguide propagation loss of 0.94dB and 1.41dB respectively. This in turn makes the insertion loss $IL^{dB}$ value (that includes the waveguide propagation loss in addition to some other loss parameters [38]) to change for each data packet transfer in the PNoC. This observation opens new opportunities for dynamically changing for every packet transfer either the $P_{Laser}$ value or the utilization (in terms of $PP^{dB}$ and/or $N_\lambda$) of the fixed design-time $P_{Laser}$ value.

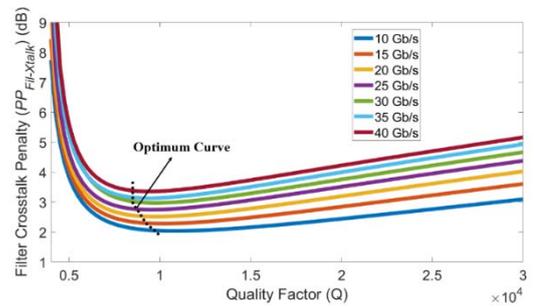

### III. Fig. 2: Filter crosstalk penalty ($PP^{Fil}_{Xtalk}$) as a function of Quality Factor (*Q*) for various values of signal bitrate (*BR*). Evaluation is done using Eq. (2) and Eq. (3) for 0.37nm wavelength spacing and $N_\lambda$=55 at 1550nm operating wavelength. Related Work and Motivation

Because of the high insertion loss and power penalty in the constituent photonic links [13][15][38], the state-of-the-art PNoC architectures require a non-trivial amount of optical power from their laser source. The high optical power overheads from the laser source can offset the high aggregated datarate, low packet frame delay, and optical power efficiency advantages of PNoCs. Therefore, it is imperative to innovate new techniques that can reduce the optical power consumption in future PNoC architectures. Several prior works have addressed this problem, as discussed next.

### A. Prior Works on Laser Power Management

Several techniques have been proposed in prior works (e.g., [1]-[9]), that aim to reduce the optical power consumption, and hence, the power consumption of laser sources in PNoCs. To achieve the power savings, a few of these techniques (e.g., [1]-[5]) leverage the temporal and spatial variations in network traffic to opportunistically adjust the $P_{Laser}$ value (i.e., optical power extracted from laser sources) by tuning or distributing the available $N_\lambda$ in the network. These methods tend to notably reduce the power in laser sources during low network load conditions. However, if the losses encountered by optical signals in the network between the sender and receiver GIs are high, these methods would still require excessive optical power from laser sources to compensate for the high losses, even under low network load conditions. In contrast, a few other techniques focus (e.g., [6], [7]) on leveraging the inherent change in $IL^{dB}$ per packet transfer to tune the $P_{Laser}$ value (output optical power from laser sources). The amount of optical power savings achieved by these methods depends on how often the $P_{Laser}$ value can be tuned in response to the changing $IL^{dB}$. In addition, recent works [8] and [9] take a holistic approach and focus on both adapting $N_\lambda$ and leveraging the change in $IL^{dB}$ using machine learning predictors, to achieve greater savings in optical power consumption.

In addition, recent works [30] and [31] employ data approximation techniques to opportunistically trade the communication reliability for reduced laser power and/or improved performance in PNoCs. However, to gain substantial benefits, these techniques require the accuracy or reliability goals of target applications to be relaxed, which can be achieved only for a select few inherently error-tolerant applications. This constraint limits the applicability of such techniques.

### B. Motivation for Rule-Based Self-Adaptation in PNoCs

Techniques from prior work that look to dynamically adapt $N_\lambda$ in response to the changing network traffic conditions need to incorporate extra mechanisms with PNoCs to *(i)* monitor the network traffic conditions at runtime, *(ii)* distribute the available $N_\lambda$ in the network, and *(iii)* communicate the tuning decisions to the off-chip laser sources. The overheads of such extra mechanisms can offset the achieved optical power benefits. Along the same lines, among the techniques that look to leverage the change in $IL^{dB}$, [7] requires an integration of costly on-chip optical amplifiers, whereas [6] requires runtime execution of optimization heuristics. Moreover, the machine learning based self-adaptation techniques from [8] and [9] can also incur high overheads of runtime inference of the machine learning models. In addition, all these techniques do not consider the power penalty ($PP^{dB}$) as an important factor that can affect the utilization of $P_{Laser}$ in PNoCs in terms of supported $N_\lambda$. As a result, these techniques often render inviably high $N_\lambda$ values, failing to obtain the practical balance between the optical power efficiency and packet frame delay (packet transfer latency).

In contrast to these dynamic techniques from prior work, we advocate for a hybrid (static + dynamic) solution as part of our proposed *PROTEUS* framework that can achieve and maintain a balance between the optical power efficiency and performance of PNoCs. The details of our proposed *PROTEUS* framework are discussed in the next section.

## IV. PROPOSED *PROTEUS* FRAMEWORK

### A. Overview

Our proposed *PROTEUS* framework enables rule-based self-adaptation in PNoCs for dynamic management of $P_{Laser}$ and performance (in terms of packet transfer latency). *PROTEUS* includes two steps. In the first design-time step (Section IV-B), *PROTEUS* performs a search heuristic based optimization to find the optimal combination of $Q$ and $BR$ that minimizes the $PP^{dB}$ and $IL^{dB}$ values for the link. This step allows *PROTEUS* to statically reduce the $P_{Laser}$ value at the design time, compared to the techniques from prior works [7] and [3], and balance the optical power efficiency of the PNoC with its packet transfer latency. Then, during the second runtime step (Section IV-C), *PROTEUS* readjusts the $BR$, and $Q$ duplet in response to the changing $IL^{dB}$ for every packet transfer, *(i)* to ensure that the provisioned $P_{Laser}$ is always utilized as fully as possible, and *(ii)* to maintain the balance between the achieved optical power efficiency and packet frame delay (packet transfer latency) for every packet transfer.

To enable dynamic readjustments (adaptation) of $Q$, *PROTEUS* incorporates the MR modulator/filter design with adaptable $Q$ from [12], after enhancing it for a faster response. Similarly, to enable adaptation in $BR$, *PROTEUS* allows a lightweight reconfiguration of the serialization and deserialization modules in each GI to enable the scaling of photonic clock rate (that directly corresponds to signal $BR$) between the baseline value of 5GHz and four discrete upscaled values (10GHz, 15GHz, 20GHz, and 25GHz). An exhaustive search-based analysis is performed offline, to find the best combinations of $Q$ and $BR$ for all possible $IL^{dB}$ values in the PNoC. From this offline analysis, simple rules are derived about what should be the change in the control parameters (e.g., reconfiguration parameters that control the dynamic clock rate scaling) to adapt $BR$ and $Q$ combination for each transferred packet, as $IL^{dB}$ changes for each packet transfer as discussed in Section II-D. These rules (i.e., new control parameter values) are stored in a lookup table at every GI of the PNoC, which *PROTEUS* refers to at runtime before each packet transfer to enable adaptation of $Q$ and $BR$.

### B. Search Heuristic Based Design-Time Optimization

In a PNoC, $IL^{dB}$ varies for different sender-receiver pairs, as illustrated in Fig. 1 (Section -D). We model all unique $IL^{dB}$ values that a photonic packet can experience across all possible sender-receiver combinations. For our PNoC in Fig. 1, the best-case $IL^{dB}$ is 0.47dB and the worst-case $IL^{dB}$ is 10dB. *Note that we consider only the waveguide propagation loss as $IL^{dB}$ for our analysis presented in this section*. Prior works [7] (henceforth identified as OPA) and [3] (identified as ABM), with which we compare our *PROTEUS* framework, do not consider the impact of $PP^{dB}$ on $P_{Laser}$ utilization, as inferred from the fact that the assumed $Q$ or $BR$ values are not reported in [7] and [3]. As a result, OPA and ABM assume inviably high value of $N_\lambda = 64$ that leads to $P_{Laser}$ to be greater than $P_{Max} = 20$dBm [22], for commonly used fixed values of $Q = 7000$ [34] and $BR = 10$ Gb/s [34][11]. Therefore, to make the implementations of OPA and ABM techniques viable, first, we identify the viable value of $N_\lambda$ (using Eq. (1)) for the worst-case $IL^{dB}$ of 10dB. For that, we consider $Q=7000$, $BR=10$Gb/s, $S=20$dBm [34], $P_{Laser}=P_{Max}=20$dBm, and $PP^{dB}$ as evaluated from Eq. (2)-(4). We found the maximum supported $N_\lambda$ to be 55, and we consider this as the design value for OPA and ABM. As our *PROTEUS* framework aims to achieve loss-aware power savings, we consider another loss-aware technique, i.e., OPA, as the baseline comparison in this section. Fig. 3(a) gives the packet frame delay and optical power efficiency (triangle shaped points) for different $IL^{dB}$ values (shown in different colors) for OPA. As evident, OPA reduces $P_{Laser}$ as $IL^{dB}$ decreases. As a result, the optical power efficiency values for OPA also reduce as $IL^{dB}$ decreases. However, as $N_\lambda=55$ and $BR=10$Gb/s are fixed for all $IL^{dB}$ cases for OPA, all $IL^{dB}$ cases achieve the same packet frame delay (Fig. 3(a)).

From these results for OPA, the goal of *PROTEUS* framework becomes to statically reduce the required $P_{Laser}$ to a value below 20dBm that can support the unchanged aggregated data-rate ($N_\lambda \times BR = 55 \times 10$Gb/s$=550$Gb/s) for all possible $IL^{dB}$ cases. Intuitively, if a $P_{Laser}$ value that is less than 20dBm can support $N_\lambda=55$ for the worst-case $IL^{dB}$ of 10dB, then that $P_{Laser}$ value can support $N_\lambda=55$ for all other $IL^{dB}$ values lower than 10dB as well. To find such $P_{Laser}$ value, *PROTEUS* aims to reduce $PP^{dB}$ for the worst-case $IL^{dB} = 10$dB, by optimizing $Q$ for the given $BR = 10$Gb/s (unchanged compared to OPA), using a search heuristic. The search heuristic takes 28 different $Q$ values (i.e., from 5000 to 12000 with step increment of 250) and finds $Q=9750$ to provide minimal $PP^{dB}$ for $BR = 10$Gbps, which corroborates with Fig. 2 where the optimum curve for $BR=10$Gbps falls at the same value of $Q = 9750$. Thus, at $Q = 9750$ we have the least $PP^{dB}$, which gives us the opportunity to statically reduce $P_{Laser}$.

### C. Impact of Varying Q and BR

From Section IV-B, intuitively any $IL^{dB}$ that is less than the worst-case value of 10dB should require less than $P_{Laser}=16$dBm. But *PROTEUS* keeps $P_{Laser}$ to be fixed at 16dBm for each packet transfer, irrespective of $IL^{dB}$. This provides an opportunity to increase $BR$ for smaller $IL^{dB}$ values, by allowing the accommodation of a larger $PP^{dB}$ value to fully utilize the provisioned $P_{Laser}$ of 16dBm. For fully utilizing the provisioned $P_{Laser}$ for different $IL^{dB}$ values, *PROTEUS* adaptively varies $Q$ and $BR$ for different $IL^{dB}$ value (i.e., for each different packet). For that, *PROTEUS* uses the offline search heuristic to find the optimal values of $Q$, $BR$ that provides the minimum positive value of $e = (P_{Laser} - IL^{dB}$

– $PP^{dB}$ – $10log(N_\lambda)$ – $S$) (derived from Eq. (1)), as the minimum value of $e$ means that $P_{Laser}$ is fully utilized for that $BR$ and $Q$ combination. Such optimal $BR$ and $Q$ values are found for each possible $IL^{dB}$ value in the PNoC. As inputs to the search heuristic, we use the same values of $Q$ as used in Section IV-B, whereas we limit $BR$ to only four discrete values of 10Gbps, 15Gbps, 20Gbps, and 25Gbps to enable a viable $BR$ adaptation control mechanism as discussed in Section V-A. From Fig. 3(a), as the $IL^{dB}$ values reduce from 10dB, the optimal $Q$ and $BR$ values for *PROTEUS* change, yielding increasingly better (lower) frame delay and optical power efficiency values.

To understand the reason behind that, consider Fig. 3(b) that plots the breakdown of $P_{Laser}$ utilization and aggregated datarate values for *OPA* and *PROTEUS* for various insertion loss ($IL^{dB}$) values. In Fig. 3(b), $P_{Laser}$ decreases for OPA as the insertion loss ($IL^{dB}$) decreases. In contrast, for *PROTEUS*, $P_{Laser}$ remains constant for all insertion loss values. However, the detector sensitivity increases as the insertion loss increases. This is because, the detector sensitivity typically increases with the increase in $BR$ [34], and from Fig. 3(a), $BR$ increases as the insertion loss ($IL^{dB}$) decreases for *PROTEUS* (circular points). Despite this increase in $BR$ with the decrease in insertion loss for *PROTEUS*, the utilization of $P_{Laser}$ for $PP^{dB}$ for *PROTEUS* remains at the minimum possible value for all insertion loss values. This contrasts with what happen for OPA (Fig. 3(b)). Such minimization of $PP^{dB}$ for all insertion loss values allows for larger $BR$ values at smaller insertion loss values for *PROTEUS*, yielding greater aggregated datarate (green columns in Fig. 3(b)) for *PROTEUS* for smaller insertion loss values. Thus, dynamic adaptation of $BR$ and $Q$ with changing insertion loss values for each packet transfer allows *PROTEUS* to opportunistically improve the frame delay and optical power efficiency for different packet transfers.

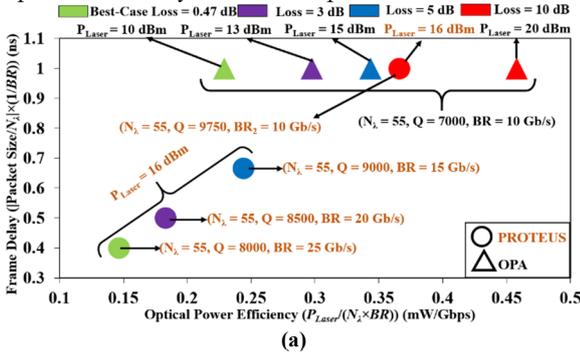

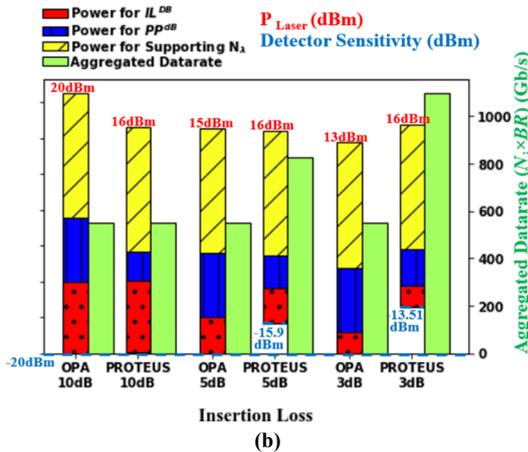

Fig 3: (a) Frame Delay and Optical Power Efficiency for different insertion loss values (indicated by different colors) for *OPA* and *PROTEUS* (indicated with different shapes); (b) Utilization of $P_{Laser}$ and Aggregated Datarate for *OPA* and *PROTEUS* across different insertion loss values.

## V. IMPLEMENTATION OF PROTEUS FRAMEWORK

Our proposed *PROTEUS* framework uses the offline search heuristic analysis described in Section IV-C to find the optimal combination of $BR$ and $Q$ for different $IL^{dB}$ values. Using this offline information, *PROTEUS* dynamically adapts $Q$ and $BR$ to the optimum values to co-optimize optical power efficiency and frame latency, for each photonic packet transfer. *PROTEUS* incorporates a lookup table at each GI, which stores the rules in terms of the required control parameter values for adapting $Q$ and $BR$. We propose to adapt $Q$ by incorporating an $MR$ modulator/filter design from [12], and adapt $BR$ by implementing a light-weight reconfiguration of the serialization and deserialization modules at each GI. We discuss the operation of these adaptive designs and their incurred overheads in the next subsections. We also discuss how we derive the rules required to enable the lookup table based adaptation.

### A. Dynamic Adaptation of BR

To enable dynamic reconfiguration of the $BR,$ we propose to use reconfigurable serializer and deserializer modules at each GI, as illustrated in Fig. 4. In the design shown in Fig. 4, at each GI, the clock distribution H-tree (implementation of which is explained in Section V-C) supplies a discrete set of upscaled clocks rate (10GHz, 15GHz, 20GHz and 25GHz). Each of these upscaled clock rates corresponds to the specific $BR$, e.g., the clock rate of 10GHz corresponds to $BR$ of 10Gb/s. In Fig. 4, each GI has multiple copies of both serializer and deserializer units, with each copy enabling the clock rate scaling between the baseline rate of 5GHz (not shown in Fig. 4) and a specific upscaled rate. For example, the '5GHz to 10GHz' ('10GHz to 5GHz') serializer (deserializer) unit enables clock-rate scaling from (to) the baseline value of 5GHz to (from) the upscaled value of 10GHz.

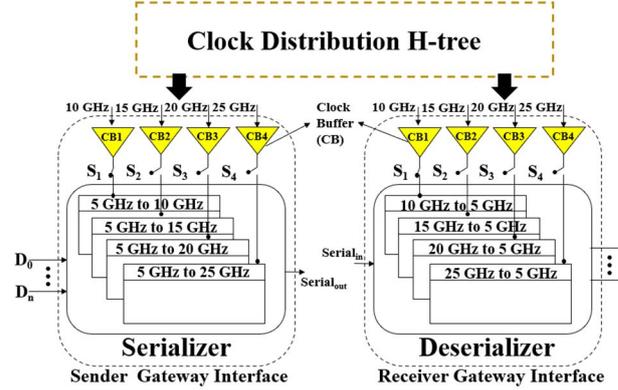

**Fig. 4:** The schematic of the reconfigurable serializer and deserializer units at the sender and receiver gateway interfaces. These units along with the upscaled clock rates provided from the clock distribution H-tree and switches $S_1$, $S_2$, $S_3$ and $S_4$ enable dynamic adaptation $BR$.

The selection of the serializer and deserializer units to be used for transmission of a photonic packet is controlled by the switches $S_1$, $S_2$, $S_3$ and $S_4$. The switches $S_1$ to $S_4$ are also used to gate the upscaled clock signals, so that the idle serializer and deserializer units can be turned off. For instance, in Fig. 4, the sender GI can serialize the input data bits ($D_0$ to $D_n$) of a packet with $BR$=10Gb/s by configuring the switches $S_1, S_2, S_3,$ and $S_4$ to '1', '0', '0', and '0' states respectively, which means that switch $S_1$ is closed and switches $S_2$ to $S_4$ are open. These states of the switches can be collectively represented with the switch-state vector $S_1S_2S_3S_4$ = '1000'. This '1000' switch-state vector basically selects the '5GHz to 10GHz' serializer unit at the sender GI and the '10GHz to 5GHz' deserializer unit at the receiver GI. It also gates the remaining three serializer and deserializer modules at the sender-receiver GIs to the power down mode. Thus, *PROTEUS* can use this switch-state vector $S_1S_2S_3S_4$ as the control parameter before each packet transfer at the sender and receiver GIs involved with the packet transfer, to select the appropriate serializer-deserializer pair and to consequently control the $BR$ for the packet transfer. The overheads of this $BR$ control mechanism are discussed next.

*1) Area and Power Overhead Analysis:* The dynamic adaptation of BR incurs overhead for the generation and distribution of various upscaled clock signals. In addition, the dynamic power overhead of the serializer and deserializer modules change with the selection of the upscaled BR. We consider the power values for the serializer and deserializer units from [27] for the 45nm CMOS SOI platform, and scale them for different upscaled BR values. Accordingly, the serializer modules corresponding to the upscaled BR values of 10Gb/s, 15Gb/s, 20Gb/s, and 25Gb/s, respectively, consume 1.4mW, 2.4mW, 3.3mW, and 4.2mW power. From [19], the deserializer units also have approximately the same power values as the serializer units. Moreover, we consider the power and area consumption of the clock generator per upscaled clock rate to be 0.5mW and 180μm$^2$. Further, the clock distribution H-tree also incurs similar area (for the required clock buffers across the H-tree network) and power overheads of 0.504mW and 320μm$^2$ per upscaled clock rate [20]. In addition, we assume that the serializer and deserializer units can be woken-up from the power-down mode in ~200ps, which we think is the reasonable value as the critical path for these units can be reasonably short [33]. time for the We include these power overhead values in our system-level simulations in Section VI.

### B. Dynamic Tuning of Q

To enable dynamic tuning of *Q*, we extend the two-point coupled MZI-based MR modulator design from [12] for a faster response. Fig 5(a) and 5(b), respectively, show our utilized MR modulator and MR filter designs. In these designs, the regular coupling waveguide, which generally supports the input and through ports of the MR device, is extended to have a long coupler arm that couples with the MR at two points C1 and C2. In the original design from [12], this coupler arm is integrated with a microheater that can thermo-optically change the coupler optical path-length $l_1$ with respect to the MR optical path length $l_2$ to modulate the coefficients of light coupling at points C1 and C2, which in turn results in the modulation of quality factor (*Q*) for the MR. Using this method, a wide range of *Q* tuning has been demonstrated in [12]. However, the use of microheater results in a very slow response time for tuning *Q* (in the order of milliseconds). Therefore, to improve the response time, we embed a reverse-biased PN-junction based phase-shifter in the coupler arm, instead of the heater based approach in [12]. From Fig. 5(a), by changing the reverse bias voltage $V_R$ across the PN-junction, the depletion layer width can be changed in the PN-junction to change the effective index of the coupler arm, which in turn can change the optical path length $l_1$ of the coupler arm, resulting in the change in the coupling coefficients and *Q* of the MR. Thus, *PROTEUS* can use the applied reverse-bias voltage across the coupler arm as the control parameter to tune the *Q* values for individual MR modulators and filters in the PNoC.

*1) Power Overhead and Response Time Analysis:*

We model our designed MRs, along with the PN-junction based phase-shifter in the coupler arms of the MRs, using the phase-shifter model given as part of the open-source modeling framework [35]. In our model, we use the nominal carrier concentration values for the P+, N+, P++, and N++ doping regions and PN-junction dimensions from [12]. From our modeling, we find that the *Q* of our designed MRs can be adapted with the response time to be in the range of ~20-30ps.

In addition, the adaptation of Q incurs Q-tuning power overhead. Fig. 6 gives the variation of *Q* with respect to the *Q-tuning power*, which is associated with $V_R$ across the PN-junction. From Fig. 6, tuning of *Q* values over a wide range can be achieved. From the figure, the highest *Q-tuning power* value is 6.1μW per MR. This value translates into total 0.03W power overhead, if the *Q* values for all 6457 MRs in our considered enhanced Flexishare PNoC archittcure [17] are tuned.

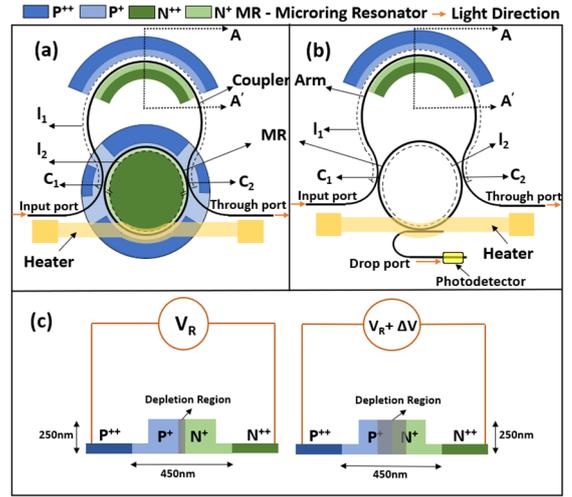

**Fig. 5:** (a) Two-point coupler arm based MR modulator; (b) Two-point coupler arm based MR filter; (c) Cross-sectional view along AA′ of the PN-junction embedded in the coupler arms of the MRs.

Further, Fig. 6 also captures the dependency of the change in extinction ratio with the change in *Q*. This dependency results in the power penalty in MR modulators due to the limited extinction ratio of modulated signals. This power penalty can be modeled using the first term in Eq. (2). For that, we evaluate *r* from the extinction ratio value obtained from Fig. 6. For example, the horizonal brown line shown in Fig. 6 corresponds to *Q*=6000 and extinction ratio = 17.5dB, which in turn corresponds to $r = 10^{(17.5/10)} \approx 56.2$. Using this *r* value in the first term of Eq. (2) yields the power penalty of 0.154dB. We evaluate this power penalty as part of our offline search heuristic described in Section IV-C. Therefore, our selected *Q* and *BR* values for different $IL^{dB}$ values (Section IV-C) already reflect this power penalty overhead.

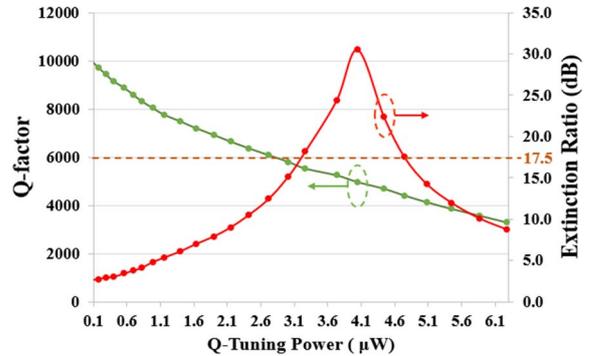

**Fig. 6:** Variation of *Q* (Q-factor) and extinction ratio in our considered MR designs from Fig. 5 with applied Q-Tuning Power.

### C. Putting All Together with Rule-Based Lookup

Fig. 7 shows the schematic implementation of our *PROTEUS* framework. From the figure, the upscaled clock signals required for *BR* adaptation (not shown in Fig. 7) are generated in the centralized clock generator, and then these clock signals are delivered to the individual GIs in the PNoC through the clock distribution H-tree. In addition, each GI in the PNoC uses an SRAM-based lookup table to stores the control parameters (i.e., the switch-state vectors $S_1S_2S_3S_4$ for *BR* and $V_R$ values for *Q*) that enable the adaptation of *BR* and *Q* for every packet transfer. Every entry in the lookup table is indexed using an ID that identifies the sender-receiver GI pair to be involved for the packet transfer associated with the entry. Before each packet transfer, the associated sender and receiver GIs access the control parameters from the lookup table and adapt the *BR* and *Q* accordingly, all during the arbitration and receiver selection phases [25] that are required for successful transfers of data packets over the crossbar based PNoCs (e.g., [17], [25]).

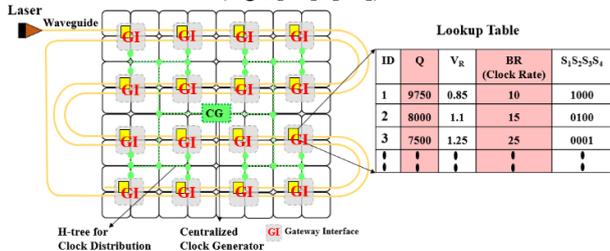

**Fig. 7:** Schematic implementation of our *PROTEUS* framework on the enhanced Flexishare PNoC architecture from [25].

*1) Overheads of Rule-Based Lookup*

The access latency of lookup table indexing is evaluated to be ~40ps, using CACTI 7.0 [16] based modeling and analysis. This latency, when added to the Q-adaptation response time of ~20-30ps and the wake-up time for the serializer-deserializer units of ~200ps, gives the total latency for Q and BR adaptation to be ~270ps, which is about half the typical processing core operating clock period of 500ps (i.e., 2GHz clock rate). Moreover, each lookup table has 64 entries (corresponding to 64 sender-receiver GI pairs) of 24-bits each, to support the storing of sender-reciever *IDs*, $V_R$ values and $S_1S_2S_3S_4$ vectors. The total area overhead of all lookup tables in the PNoC is 0.09mm$^2$.

## VI. EVALUATION

### A. Evaluation Setup

For evaluating our *PROTEUS* framework, we simulate a 256-core system with a PNoC that has 32 GIs and 32 clusters, with each cluster having 8 cores. We targeted a 22nm process node and 5 GHz clock frequency for the 256-core system. We consider the recently proposed variant [25] of the well-known Flexishare PNoC architecture [17], which employs the overlapped concurrent token stream arbitration method. Fig. 1 shows the physical-layer schematic of the scaled down version (i.e., 64 cores, 16 clusters, 16 GIs, 4 cores per cluster) of our considered PNoC. Our considered PNoC architecture implements intra-cluster communication in the electronic domain and inter-cluster communication in the photonic domain, as done in the PNoC from [26]. Our considered PNoC uses 32 multiple-writer-multiple-reader (MWMR) type of crossbar waveguides, with each waveguide employing total 55 DWDM photonic signals (i.e., $N_\lambda$ = 55). We consider a packet size (*PS*) of 512 bits, therefore, the frame delay for our PNoC becomes $\{|(PS/N_\lambda)| \times (1/BR)\} = 10/BR$. We modeled and simulated the architectures at cycle-accurate granularity with a SystemC-based in-house NOC simulator. We used real world traffic from applications in the PARSEC benchmark suite [27]. The traces of PARSEC benchmark applications were generated from gem5 full-system simulations, and then were fed into our NoC simulator. We adequately warmed up our Gem5 simulations to consequently extract the traces from the regions-of-interest (ROIs) [14] of the applications.

To compute laser power, we considered the values listed in Table   for calculating the total optical power coupled to the PNoC chip. Then, we considered the wall-plug efficiency of 10% to evaluate the electrical input power in the off-chip laser source (referred to as electrical laser power). In addition to the electrical laser power, we also evaluated the average packet latency and aggregated energy-per-bit (EPB) values. We evaluate aggregate EPB as the sum of electrical laser EPB, thermal tuning EPB, and overhead EPB. To evaluate EPB, we divide average power value with the average throughput of the PNoC, e.g., to evaluate electrical laser EPB, we divide electrical laser power with the average throughput, and vice versa. We take our overhead power values from Section   , and thermal tuning power from Table 1.

**TABLE 1. VARIOUS LOSS AND POWER PARAMETERS**

| Parameter | Value |
|---|---|
| Laser wall-plug efficiency | 10% [34] |
| Sensitivity at 10Gb/s | -20dBm [34] |
| Waveguide Insertion Loss | 0.54dB/cm [24] |
| Waveguide Bending Loss | 0.005 dB/90$^0$ [38] |
| Splitter Loss | 0.5dB [38] |
| Coupler Loss | 2dB [38] |
| Free Spectral Range (FSR) | 20 nm [36] |
| Thermal tuning power | 800 μW/nm [37] |

We compared *PROTEUS* with two dynamic laser power (LP) management techniques from prior work: Adaptive Bandwidth Management technique (ABM) from [3], and On-chip Semiconductor Amplifier (OPA) based technique from [7]. ABM performs a weighted time-division multiplexing of the photonic network bandwidth and leverages the temporal fluctuations in network bandwidth to opportunistically save LP. ABM is designed to perform LP management in MWMR waveguides [3]. On the other hand, OPA uses on-chip semiconductor amplifiers to achieve traffic-independent and loss-aware savings in LP. We consider Flexishare with ABM as our base case for comparison. For comparison with ABM, it is necessary to enable weighted time division multiplexing of the network bandwidth in the Flexishare PNoC. Therefore, we enhanced the Flexishare PNoC [17] with the overlapped concurrent token stream arbitration method from [25], to enable weighted time-division multiplexing of the network bandwidth. We analyzed the power dissipation, average packet latency and aggregate EPB for OPA, ABM and *PROTEUS*, when these frameworks were integrated with our considered enhanced Flexishare PNoC architecture.

### B. Comparative Analysis Results

Fig. 8 presents total power dissipation (sum of electrical laser power, thermal tuning power, and overhead power) results for ABM, OPA and *PROTEUS*. ABM does not have any power overheads involved [3], whereas OPA has the power overhead of tuning OPAs [7] and *PROTEUS* has the overhead of adapting *Q* and *BR*. Despite of this fact, the total power consumption for *PROTEUS* is less than ABM by 17.89%. This is because the static reduction in optical $P_{Laser}$ to 16dBm for *PROTEUS* turns out to be significant than the dynamic and traffic-dependent reduction in optical $P_{Laser}$ for ABM, which in turn reduces the electrical laser power for *PROTEUS* by 24.5%, contributing to the reduction in total power consumption. In contrast, *PROTEUS* consumes 5.13% more total power than OPA, despite OPA consuming significantly more overhead power than *PROTEUS*. This is because the optical $P_{Laser}$ is modulated to its minimum required value for every packet transfer in OPA, which proves to be better than the static reduction in $P_{Laser}$ achieved by *PROTEUS*, resulting in less total power consumption for OPA than *PROTEUS*. Nevertheless, *PROTEUS* achieves better average latency and EPB results, as discussed next.

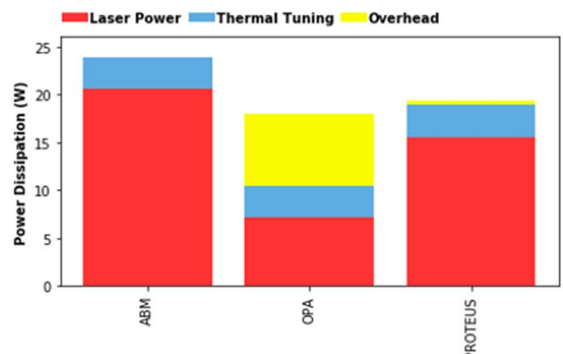

**Fig. 8:** Total power (electrical laser, thermal tuning, and overhead power) dissipation results for the ABM, OPA, and *PROTEUS* enabled variants of our considered enhanced Flexishare PNoC architecture. Overhead is for adapting Q and BR.

Fig. 9 shows the average packet latency results, with all values normalized with respect to the ABM technique. As evident from Fig. 8, it can be observed that on average, *PROTEUS* achieves 31% and 21.5% lower latency than ABM and OPA, respectively. The lower latency for *PROTEUS* is due to the dynamic adaptation of Q and BR, which decreases the $PP^{dB}$ to increase the aggregated datarate and reduce the frame delay, resulting in reduced latency. In contrast, ABM and OPA techniques do not aim to reduce average packet latency at all. Furthermore, ABM experiences higher latency compared to OPA, as ABM incurs additional latency for switching ON and OFF the off-chip laser sources [3].

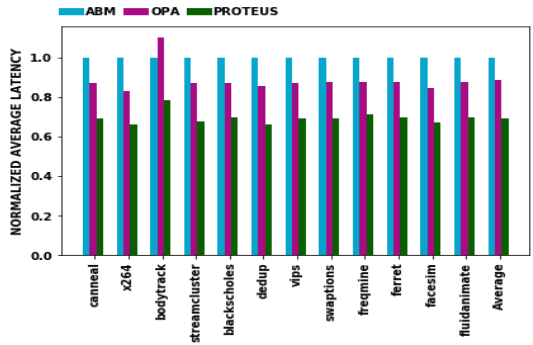

**Fig. 9: Normalized average latency for the ABM, OPA, and *PROTEUS* enabled variants of our considered enhanced Flexishare PNoC architecture. Results are normalized with respect to the ABM technique.**

Fig. 10 gives aggregate EPB (sum of electrical laser EPB, thermal tuning EPB, and overhead EPB) results for ABM, OPA, and *PROTEUS*. *PROTEUS* consumes 20% and 13.6% less aggregate EPB than ABM and OPA, respectively. As seen earlier, *PROTEUS* has the least average latency, which yields the highest throughput, resulting in the least EPB, compared to ABM and OPA. Moreover, as the average latency and total power for ABM are higher than OPA, ABM has greater aggregate EPB than OPA. Thus, our proposed *PROTEUS* framework is able to strike a balance between the total power consumption and performance (in terms of average packet latency) of the PNoC, and therefore, it can achieve more energy-efficiency in terms of energy-per-bit.

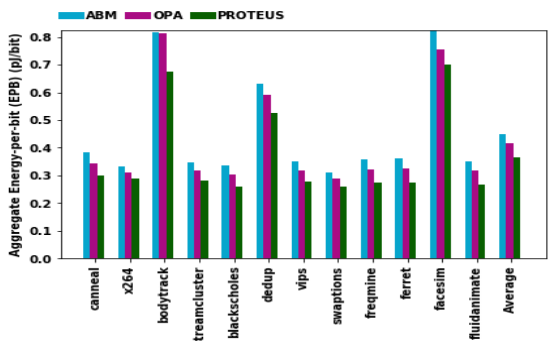

**Fig. 10: Aggregate energy-per-bit (EPB) results for the ABM, OPA, and *PROTEUS* enabled variants of our considered enhanced Flexishare PNoC architecture.**

## VII. CONCLUSIONS

This paper presented an insertion loss aware framework *PROTEUS* that enables rule-based dynamic adaptation of the key photonic link configuration parameters, such as Q-factor of microrings and bitrate of photonic data signals, to statically reduce the laser power consumption and opportunistically improve the packet transfer latency in PNoCs. *PROTEUS* exploits the dependence of BER power penalty in PNoCs on Q-factor and bitrate to balance the reduction in laser power consumption in PNoCs with the achieved aggregated datarate and packet latency. Evaluation with PARSEC benchmarks shows that the *PROTEUS* framework can achieve up to 24.5% less laser power consumption, up to 31% less average packet latency, and up to 20% less energy-per-bit, compared to two other laser power management techniques from prior work. Thus, *PROTEUS* represents an attractive solution for co-optimizing the laser power consumption and performance of emerging PNoCs.